\documentclass[aps,prr,twocolumn,floatfix,showpacs,citeautoscript,longbibliography,superscriptaddress]{revtex4-2}

\usepackage{graphicx}
\usepackage{dcolumn}
\usepackage{bm}
\usepackage{bbold}
\usepackage{mathtools}
\usepackage{color}
\usepackage{transparent}
\usepackage{dsfont}
\usepackage{amsmath}
\usepackage[colorlinks=true,citecolor=blue]{hyperref}
\newcommand{\revision}[1]{{{#1}}}

\begin{document}
\title{Photon emission statistics of a driven microwave cavity}
\author{Pedro Portugal}
\affiliation{Department of Applied Physics, Aalto University, FI-00076 Aalto, Finland}
\author{Fredrik Brange}
\affiliation{Department of Applied Physics, Aalto University, FI-00076 Aalto, Finland}
\author{Kalle S. U. Kansanen}
\affiliation{Department of Applied Physics, Aalto University, FI-00076 Aalto, Finland}
\affiliation{Department of Physics and NanoLund, Lund University, Box 188, SE-22100 Lund, Sweden}
\author{Peter Samuelsson}
\affiliation{Department of Physics and NanoLund, Lund University, Box 188, SE-22100 Lund, Sweden}
\author{Christian Flindt}
\affiliation{Department of Applied Physics, Aalto University, FI-00076 Aalto, Finland}
\begin{abstract}
Recent experimental advances have made it possible to detect individual quantum jumps in open quantum systems, such as the tunneling of single electrons in nanoscale conductors or the emission of photons from non-classical light sources. Here, we investigate theoretically the  statistics of photons emitted from a microwave cavity that is driven resonantly by an external field. We focus on the differences between a parametric and a coherent drive, which either squeezes or displaces the cavity field.  We employ a Lindblad master equation dressed with counting fields to obtain the generating function of the photon emission statistics using a theoretical framework based on Gaussian states. We then compare the distribution of photon waiting times for the two drives as well as the $g^{(2)}$-functions of the outgoing light, and we identify important differences between these observables. In the long-time limit, we analyze the factorial cumulants of the photon emission statistics and the large-deviation statistics of the emission currents, which are markedly different for the two drives. Our theoretical framework can readily be extended to more complicated systems, for instance, with several coupled microwave cavities, and our predictions may be tested in future experiments. 
\end{abstract}

\maketitle
\section{Introduction}

The control of single photons in the microwave regime is gathering increasing interest as recent quantum technologies have paved the way for accurate emitters and detectors of microwave photons~\cite{Gu2017,Houck2007,Gleyzes2007,Johnson2010,Bozyigit2010,Sathyamoorthy2014,Kono2018,Royer2018,pizzi2022light}. Detectors may be based on double quantum dots in which the absorption of a photon causes the inelastic tunneling of an electron from one quantum dot to the other~\cite{Khan2021,Zenelaj2022,Haldar2023}. Calorimetric schemes have also been developed, whereby the temperature of a mesoscopic reservoir abruptly increases upon the absorption of a photon~\cite{Brange2018,Karimi2020,Karimi2020b}. For microwave photons, unlike their optical counterparts, thermal effects are important, since the photon energies can be comparable to the temperature in sub-Kelvin experiments. In this context, the heat carried by photons in microwave cavities \cite{Bergenfeldt2014} or electrical circuits \cite{Golubev2015} has been investigated, and, more generally, the statistics of photon transfers has become an important topic in quantum thermodynamics~\cite{Vinjanampathy2016,Anders2017}, for instance, in connection to thermodynamic uncertainty relations for open quantum systems~\cite{Menczel2020,menczel2021thermodynamic}.

Theoretically, the photon counting statistics of parametric amplifiers at zero temperature has been investigated~\cite{Vyas1989} as well as the photon-number fluctuations in driven resonators~\cite{PhysRevA.75.042302,PhysRevA.84.043824,PhysRevLett.116.013603}. In addition, the statistics of photons exchanged between a thermal microwave cavity and its environment has been explored~\cite{brange2019photon} together with the number of photons that are transferred between a microwave cavity and an external driving field~\cite{menczel2021thermodynamic}.  Typically, the photons are non-interacting, such that the problem is quadratic in the creation and annihilation operators, and in that case, a powerful theoretical framework based on Gaussian states can be employed~\cite{Adesso}. Such theoretical investigations are motivated by the ongoing efforts to realize efficient detectors of single microwave photons, which may soon make it possible to measure the photon counting statistics in real time.

\begin{figure}[bh!]
    \centering
    \includegraphics[width=0.45\textwidth]{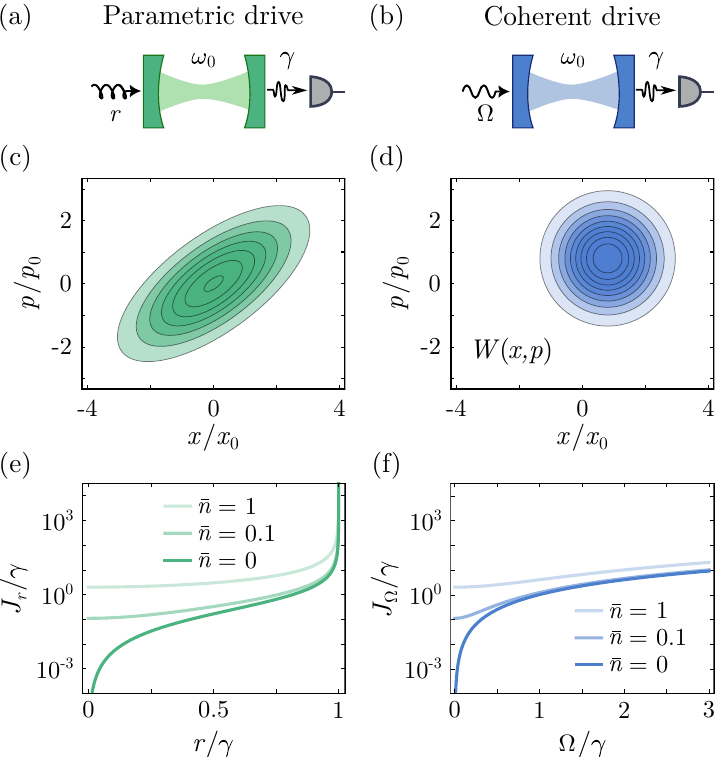}
    \caption{Photon emissions from a driven microwave cavity. (a,b) A microwave cavity is driven at its resonance frequency~$\omega_0$, and photons are emitted into a thermal environment at the rate $\gamma$. We compare the parametric drive in~(a) with strength $r$ and phase $\phi_r$ to the coherent drive in (b) with Rabi coupling~$\Omega$ and phase $\phi_\Omega$. (c,d) Phase-space representations of the stationary states, which at zero temperature are a squeezed state (c) and a coherent state (d), respectively. Here, the usual oscillator length is denoted by $x_0$, and the corresponding momentum is $p_0=\hbar/x_0$. The parameters for the parametric drive are $r = 0.7\gamma$ and $\phi_r = 0$, and for the coherent drive they are $\Omega = 0.7\gamma$ and $\phi_\Omega = \pi/4$. (e,f) Average emission currents for different temperatures given by~$\bar n$.}
    \label{fig:fig1}
\end{figure} 

Here, we investigate the photon emission statistics from a microwave cavity that is either parametrically driven or coherently driven as illustrated in Fig.~\ref{fig:fig1}(a,b). Due to the external drive, the cavity field is either squeezed or displaced as shown in Fig.~\ref{fig:fig1}(c,d). The photon emission statistics is encoded in a generating function, which we obtain from a Lindblad master equation dressed with counting fields. Technically, we solve the master equation using a theoretical framework based on Gaussian states, which leads to a Riccati equation whose solution eventually yields the generating function. Here, we note that related approaches were recently developed  both for bosons~\cite{kerremans2022probabilistically,landi2023current,kim2023} and fermions~\cite{brenes2022particle}. With the generating function at hand, we can analyze the differences between the drives in the time domain by evaluating the distribution of photon waiting times and the $g^{(2)}$-correlation functions. We also consider the number of photons that are emitted over a long time interval by analyzing the factorial cumulants of the emission current together with its large-deviation statistics. Our predictions may be tested in future experiments on driven microwave cavities combined with efficient photon detectors.

The paper is organized as follows. In Sec.~\ref{ch:hamiltonian}, we introduce the microwave cavity together with the parametric drive and the coherent drive that we investigate. We describe the Hamiltonian of the driven cavity together with the Lindblad master equation that accounts for the exchange of photons with a thermal environment. In Sec.~\ref{ch:MGF}, we find the generating function of the photon emission statistics, which forms the backbone of the paper, and which we use to obtain all other results in the following sections. In Sec.~\ref{ch:waitingtimes}, we evaluate the distribution of waiting times between subsequent photon emissions and identify important differences between the two drives. In Sec.~\ref{ch:g2fun}, we consider the $g^{(2)}$-function of the outgoing photons, which we use to show that the photon emissions do not constitute a renewal process with uncorrelated waiting times. In Sec.~\ref{ch:longtime}, we turn to the long-time limit of the photon current, and we find that all factorial cumulants are positive, which seems to be a typical property of non-interacting bosons. Finally, in Sec.~\ref{ch:largedev}, we evaluate the large-deviation statistics of the photon current and find that they are markedly different for the two drives. In Sec.~\ref{sec:conc}, we present our conclusions together with an outlook on possible developments for the future. Several technical details are deferred to the Appendices, including a brief discussion of Gaussian states and our derivation of the generating functions, which is based on the solution of a Riccati equation.

\section{Driven microwave cavity}\label{ch:hamiltonian}
We consider a microwave cavity that is driven by an external field, and we focus on a single cavity mode with frequency $\omega_0$. In particular, we are interested in comparing the photon emission statistics due to different drives. On the one hand, we consider a parametric drive described by the time-dependent Hamiltonian
\begin{equation}
\hat H_r(t) = \hbar \omega_0 \hat a ^\dagger \hat a+ \hbar r {\left(\hat a^2 e^{-i2(\omega_r t-\phi_r) }+\hat{a}^{\dagger 2} e^{i2(\omega_r t-\phi_r)}\right)}/2,
\end{equation}
where $\hat a^\dagger$ and $\hat a$ are the creation and annihilation operators of photons in the cavity, and the cavity field is squeezed by an external pump field with frequency~$\omega_r$, phase~$\phi_{r}$, and nonlinear gain coefficient~$r$, see Fig.~\ref{fig:fig1}(a)~\cite{carmichael2009statistical}. On the other hand, we consider a coherent drive described by the time-dependent Hamiltonian
\begin{equation}
 \hat H_\Omega(t) = \hbar \omega_0 \hat a ^\dagger \hat a + \hbar\Omega {\left(\hat a e^{-i(\omega_\Omega t-\phi_\Omega)}+\hat{a}^\dagger e^{i(\omega_\Omega t-\phi_\Omega)}\right)}/2,
\end{equation}
 where the external microwave field has the frequency $\omega_\Omega$, phase~$\phi_{\Omega}$, and Rabi coupling $\Omega$, see Fig.~\ref{fig:fig1}(b). We focus on resonant driving, so that $\omega_{r/\Omega}=\omega_0$ in the two cases. We then switch to a frame that rotates with the frequency~$\omega_0$ and find  the time-independent Hamiltonians
\begin{equation}
\hat H_r = \hbar r (\hat a^2 e^{-i2\phi_r}+\hat{a}^{\dagger 2} e^{i2\phi_r})/2
\label{eq:hamiltonian_par}
\end{equation}
and
\begin{equation}
\hat H_\Omega = \hbar\Omega (\hat a e^{-i\phi_\Omega}+\hat{a}^\dagger e^{i\phi_\Omega})/2.
\label{eq:hamiltonian_coh}
\end{equation}
In addition, the cavity is weakly coupled to an environment, such that the density matrix of the cavity, $\hat \rho(t)$, evolves according to the Lindblad equation~\cite{breuer2002theory}
\begin{equation}\label{eq:lindblad}
     \frac{d\hat\rho(t)}{d t} = \mathcal L \hat\rho(t) = -\frac{i}{\hbar}[\hat H,\hat \rho(t)] + \frac{\gamma}{2}\mathcal D \hat \rho(t),
\end{equation}
where the Liouvillian $\mathcal{L}$ consists of two parts. The first term with the commutator describes the coherent evolution of the cavity due to the Hamiltonians in Eqs.~(\ref{eq:hamiltonian_par})~and~(\ref{eq:hamiltonian_coh}), $\hat H= \hat H_{r/\Omega}$, while
the incoherent dynamics due to the environment is governed by the dissipator
\begin{equation}
    \mathcal D\hat\rho= (\bar n+1)(2\hat a \hat\rho \hat{a}^\dagger -\{\hat{a}^\dagger \hat a,\hat\rho\})+  \bar n(2\hat{a}^\dagger \hat\rho \hat a -\{\hat a \hat{a}^\dagger,\hat\rho\}).
\end{equation}
Here, the equilibrium occupation of the cavity at the inverse temperature $\beta$ is denoted by $\bar n = 1/(\exp[\beta \hbar \omega_0]-1)$, and~$\gamma$ is the coupling between the cavity and the environment.
The first term in the dissipator describes photon emissions to the environment, while the second one corresponds to photon absorptions from the environment. We note that the parametrically        driven cavity only reaches a stationary state if the coupling is smaller than the decay rate, $r<\gamma$. If not, the system becomes unstable~\cite{carmichael2009statistical}. 

In Figs.~\ref{fig:fig1}(c) and~(d), we show the Wigner phase-space representation of the stationary state given by the equation $\mathcal L \hat \rho_s = 0$. The Wigner function is defined as
\begin{equation}
    W(x,p) = \frac{1}{\pi}\int d  q  \langle x+ q|\hat \rho | x- q\rangle e^{2 i q p/\hbar},
\end{equation}
and, for both drives, the stationary state has a Gaussian Wigner function, implying that they are Gaussian states, see Appendix~\ref{ap:gauss}. For the parametric drive, the stationary state is a squeezed thermal state, while for the coherent drive at zero temperature, it is a displaced coherent state.

\section{Photon emission statistics}\label{ch:MGF}
We are interested in the probability~$P(n,t)$ that $n$ photons have been emitted into the environment during the time span $[0,t]$. To this end, it is convenient to introduce the probability generating function
\begin{equation}
G(s,t)= \sum_{n=0}^\infty P(n,t)s^n.
\end{equation}
In addition, it will be useful to consider the factorial moment generating function, which is related to the probability generating function by a simple change of variables,
\begin{equation}
\mathcal{M}(\zeta,t) = G(\zeta+1,t)=\sum_{n=0}^\infty P(n,t)(\zeta+1)^n.
\end{equation}
By differentiating it with respect to $\zeta$ at $\zeta=0$, we obtain the factorial moments of the number of emitted photons, 
\begin{equation}
\mu_m = \partial_\zeta^m \mathcal{M}(\zeta,t)\vert_{\zeta = 0}. 
\end{equation}
The factorial cumulants follow in a similar way as
\begin{equation}
\kappa_m = \partial_\zeta^m \ln \mathcal{M}(\zeta,t)\vert_{\zeta = 0}. 
\end{equation}
Factorial moments and cumulants are useful to characterize positive discrete quantities, and the factorial cumulants are defined in such a way that only the first one is nonzero for a Poisson distribution~\cite{PhysRevB.83.075432,Kambly2013,Stegmann2015,Stegmann2016,PhysRevB.94.125433,Stegmann2017,Kleinherbers2018,10.21468/SciPostPhys.10.1.007,PhysRevB.104.125431,Kleinherbers2021,besson2021,stegmann2022,PhysRevB.105.155421}. By contrast, ordinary cumulants are defined so that only the first two cumulants are nonzero for a Gaussian distribution. Since the probability generating function and the factorial moment generating function are related by a simple change of variables, we will refer to both of them as the generating function whenever confusion can be avoided.

To obtain the photon emission statistics, we unravel the Lindblad equation with the respect to the number of emitted photons using standard techniques~\cite{Plenio1998}. The emission statistics then follow as $P(n,t)=\mathrm{tr}\{\hat \rho (n,t)\}$, where the density matrix has been resolved with respect to the number of emitted photons. In addition, by defining $\hat \rho (\zeta,t)=\sum_{n=0}^\infty \hat\rho(n,t)(\zeta+1)^n$, the generating function becomes
$\mathcal M(\zeta,t) = \mathrm{tr}\{\hat \rho (\zeta,t)\}$. Importantly, this density matrix obeys the generalized Lindblad equation~\cite{brange2019photon}
\begin{equation}\label{eq:lindbladcf}
    \frac{d\hat \rho(\zeta,t)}{d t} = \mathcal L_\zeta\hat\rho(\zeta,t)=\mathcal{L}\hat\rho(\zeta,t) + \zeta \revision{\gamma (\bar{n}+1)}\hat a \hat \rho(\zeta,t) \hat a^\dagger,
\end{equation}
where the counting field $\zeta$ now enters the Liouvillean~$\mathcal L_\zeta$. The procedure for solving Eq.~(\ref{eq:lindbladcf}) is described in Appendices~\ref{ap:gauss} and~\ref{ap:MGF} and involves the use of Gaussian states. 

For the parametric drive, we find
\begin{equation}\label{eq:MGFr}
    \mathcal{M}_r(\zeta,\tau)= \prod_{\nu=\pm \tilde r}\frac{ \sqrt{2 \xi_\nu} e^{ \tau(\xi_\nu+\nu +1) /4}}{\sqrt{e^{\tau \xi_{\nu}}(\xi_{\nu}+\chi_{\nu})+\xi_{\nu}-\chi_{\nu}}} ,
\end{equation}
where we have introduced the functions
\begin{equation}
\label{eq:xi}
 \xi_\nu = \sqrt{(1+ \nu)^2-2 \zeta  (\bar n+1) ( 2 \bar n- \nu)}
 \end{equation}
 and
\begin{equation}
\label{eq:chi}
\chi_\nu =  \left[ (1+ \nu)^2 - \zeta (\bar n+1) ( 2 \bar n-\nu) \right]/(1+ \nu).
\end{equation}
We have also defined the dimensionless coupling and time
\begin{equation}\label{eq:dimensionless}
    \begin{split}
    \tilde r &= r/\gamma,\\
    \tau &= \gamma t.
    \end{split}
\end{equation}
For $r=0$, Eq.~\eqref{eq:MGFr} reduces to the generating function of a thermal cavity that was found in Ref.~\cite{brange2019photon}. Moreover, for $\bar n=0$, we recover the generating function at zero temperature obtained in Ref.~\cite{Vyas1989}.

The generating function consists of two factors that are generating functions on their own, each corresponding to the squeezing of the cavity state along one of the two principal axes in the phase space. As a result, each factorial cumulant is a sum of two terms. For example, the dimensionless average photon current reads
\begin{equation}\label{eq:currentsr}
        \tilde J_r =\kappa_1/\tau = \frac{(\bar n+1)}{2}\left[\frac{\bar n-\tilde r/2}{1+\tilde r}+\frac{\bar n+\tilde r/2}{1-\tilde r}\right],
\end{equation}
which, without the drive, reduces to the average photon emission rate from a cavity in thermal equilibrium,
\begin{equation}
\tilde J_0 = \bar n(\bar n+1).
\label{eq:ave_eq_curr}
\end{equation}
The emission current from a cavity to a thermal reservoir can also be written as $\tilde J_r =\bar{n}_r(1+\bar{n})$, where $\bar{n}_r$ is the average number of photons in the cavity and $\bar{n}$ is given by the temperature of the reservoir. From Eq.~(\ref{eq:currentsr}), we can then identify the  number of photons in the cavity as
\begin{equation}
\bar{n}_r = \frac{\bar n+\tilde r^2/2}{1-\tilde r^2},
\end{equation}
which diverges together with the current as $\tilde r\rightarrow 1$.

For the coherently driven cavity, we find
\begin{equation}\label{eq:MGFdrive}
    \mathcal{M}_\Omega(\zeta,\tau)= \mathcal M_0(\zeta,\tau) e^{ \mathcal{C}_\Omega(\zeta,\tau) },
\end{equation}
where $\mathcal M_0(\zeta,\tau)$ is given by Eq.~(\ref{eq:MGFr}), and we have defined
\begin{equation}\label{eq:driveCGF}
        \mathcal{C}_\Omega(\zeta,\tau) =\tilde \Omega^2\frac{   \xi_0 \tau \cosh[\frac{   \xi_0 \tau}{4}]+ (4 (\xi_0 ^2-1)+   \xi_0 ^2 \tau) \sinh[\frac{   \xi_0 \tau}{4}]}{4 \bar n  \frac{\xi_0 ^3}{1-\xi_0 ^2} {\left( \cosh  [\frac{   \xi_0\tau}{4}]+\xi_0 \sinh  [\frac{   \xi_0  \tau}{4}]\right)}}
\end{equation}
together with the dimensionless coupling
\begin{equation}
    \tilde \Omega = \Omega/\gamma,
\end{equation}
while $\xi_0$ is given by Eq.~\eqref{eq:xi}. Similar to the parametric drive, we can identify two independent emission processes. The first factor of the generating function, $\mathcal{M}_0(\zeta,\tau)$, accounts for photon emissions due to thermal excitations that also occur without the drive. The second factor, $e^{ \mathcal{C}_\Omega(\zeta,\tau) }$, describes emission processes due to the drive. We note that a similar factorization has been found for the distribution of photons inside the cavity~\cite{PhysRevA.84.043824}. We also see that, for both drives, the phases, $\phi_{r/\Omega}$, are unimportant for the photon emission statistics.

\begin{figure*}
    \centering
    \includegraphics[width=0.97\textwidth]{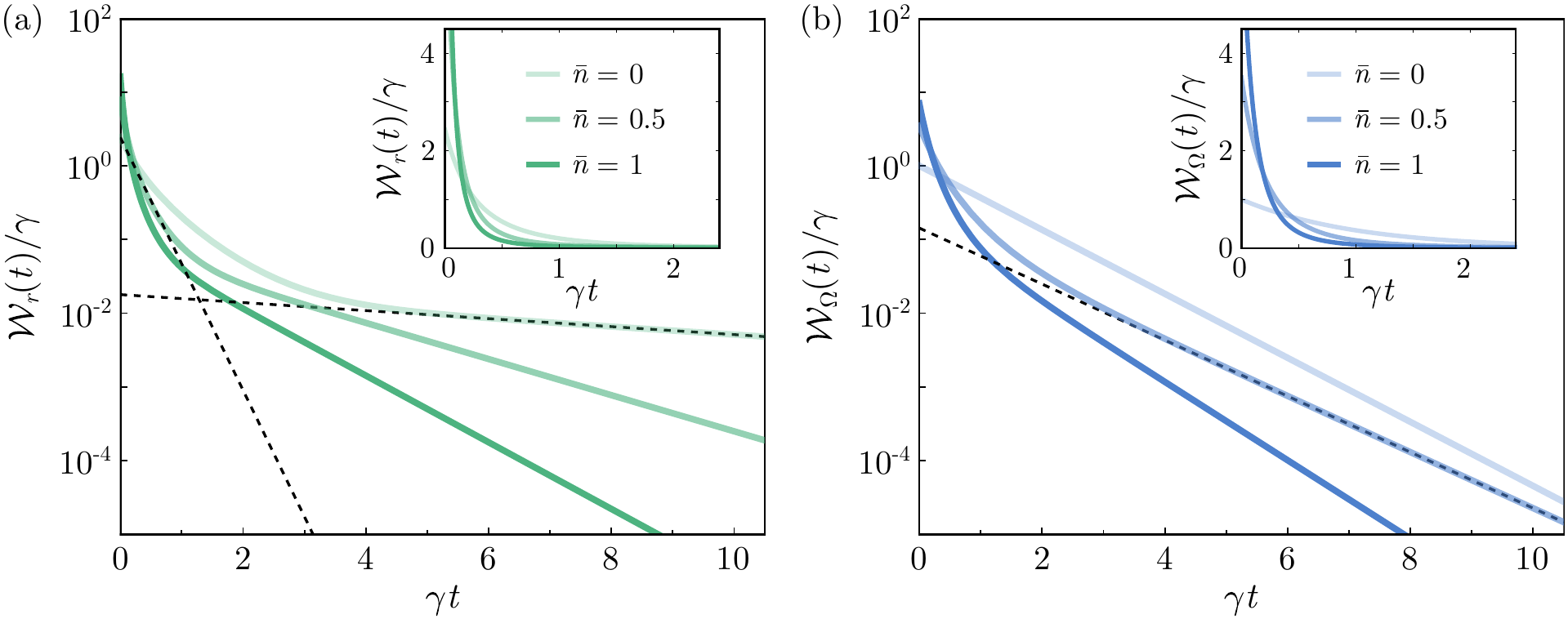}
    \caption{Waiting time distributions. (a) Waiting time distributions for the parametric drive with the coupling $r=0.75\gamma$ and different temperatures given by $\bar n$. The dashed lines are the exponential short and long time limits in Eqs.~(\ref{eq:WTDshort}) and (\ref{eq:WTDlong}). (b) Waiting time distributions for the coherent drive with $\Omega = \gamma$ and different temperatures. The dashed line indicates an exponential decay with the rate given by Eq.~(\ref{eq:longt_decay}). In both panels, the insets show the distributions on a linear scale.}
    \label{fig:wt}
\end{figure*}

From the generating function, we find the average photon current, which takes on the form
\begin{equation}
    \tilde J_{\Omega} = \tilde J_0+\tilde \Omega^2(\bar n+1),
    \label{eq:currentsOmega}
\end{equation}
where $\tilde \Omega^2 (\bar n+1)$ is a contribution from the drive that adds to the thermal part. Again, we may also identify the average number of photons in the cavity, which reads
\begin{equation}
\bar{n}_{\Omega} =\bar n+\tilde \Omega^2.
\end{equation}
As shown in Fig.~\ref{fig:fig1}(f), the temperature dependence on the average current is weak for large couplings, $\tilde \Omega$, while it becomes stronger for smaller values of $\tilde \Omega$. At low temperatures, the photon emission statistics are dominated by the contribution from the drive, and it becomes Poissonian with the rate $\tilde \Omega^2$, since $\mathcal C_\Omega(\zeta,\tau) = \tilde \Omega^2 \tau  \zeta $ for $\bar n=0$.

\section{Waiting time distributions}
\label{ch:waitingtimes}

To go beyond the information contained in the average current, we consider the distribution of waiting times between consecutive photon emissions, which we denote by~$\mathcal{W}(\tau)$~\cite{PhysRevA.39.1200,Brandes2008}. Waiting time distributions have been measured both for photon emissions \cite{PhysRevLett.112.116802} and electron tunneling \cite{PhysRevApplied.8.034019,Brangeeabe0793,Ranni2021}, and the measurements are demanding, since they require detectors with a high fidelity as no events should be missed. For stationary processes, the distribution depends only on the time difference between emissions, and it can be obtained directly from the generating function. Specifically, it can be written as~\cite{PhysRevLett.108.186806,PhysRevB.90.205429}
\begin{equation}
\label{eq:WTD}
\mathcal{W} (\tau) = \langle \tau\rangle \partial_{\tau}^2 \Pi (\tau),
\end{equation}
where $\langle \tau\rangle$ is the mean waiting time, and $\Pi(\tau)$ is the idle-time probability that no photons are emitted during a time span of duration $\tau$. The idle-time probability can be obtained from the generating function since $\Pi(\tau) = \mathcal{M}(-1,\tau)=P(n=0,\tau)$ by definition. Moreover, the mean waiting time can be related to the idle-time probability as $\langle \tau\rangle=-1/\dot{\Pi}(0)=1/\tilde J$, where we have also used that it is given by the inverse average emission current. Physically, the two time derivatives in Eq.~(\ref{eq:WTD}) can be interpreted as the detection of a photon emission at the beginning and at the end of the time interval~\cite{PhysRevB.91.195420}. 

\begin{figure*}
    \centering
    \includegraphics[width=\textwidth]{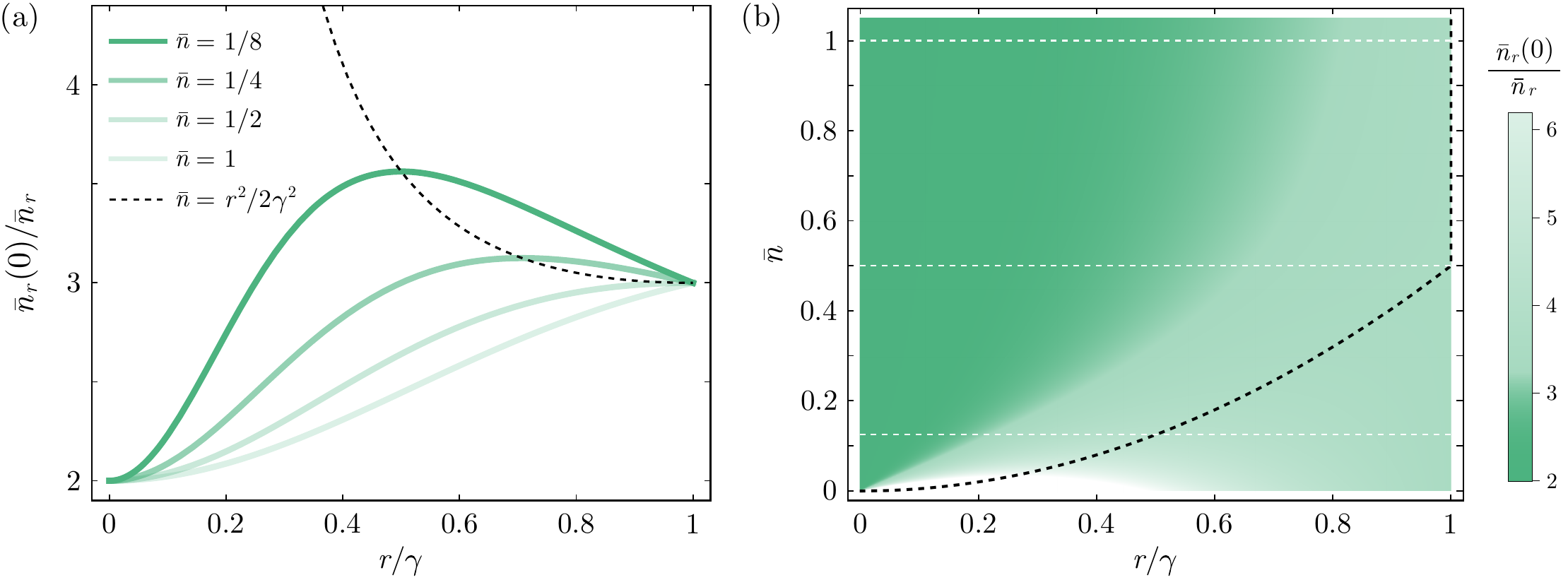}
    \caption{Photons in the cavity after a photon emission. (a) Average number of photons in the parametrically driven cavity after a photon emission as a function of the coupling and with different temperatures. The dashed line indicates the maximum value for a given coupling, which is obtained at a temperature given by the relation $\bar n = 2(r/\gamma)^2$. (b) Same results as a function of the coupling and the temperature with the maximum value for a given coupling indicated by a black dashed line. The white dashed lines correspond to each the four colored curves in panel (a).}
    \label{fig:gain}
\end{figure*}

In Fig.~\ref{fig:wt}, we show waiting time distributions for the two drives with different temperatures. At low temperatures, the waiting time distributions are clearly different for the two drives, whereas they become similar as the temperature is increased, and the thermal contribution starts to dominate. At low temperatures, the waiting time distribution for the parametric drive becomes bi-exponential with decay rates that can be determined at zero temperature. In particular, for $\bar n=0$, we find 
\begin{equation}
 \mathcal{W}_r (\tau) \simeq  \frac{1+2 \tilde r^2}{2-2 \tilde r^2}e^{-\gamma_i^{\tilde r}\tau},\ \tau\ll 1
     \label{eq:WTDshort}
\end{equation}
and
\begin{equation}\mathcal{W}_r (\tau)  \simeq \frac{\sqrt{1-\tilde r^4} {\left(\tilde r^2-\tilde r^4\right)}}{{\left(\tilde r^2+2 \sqrt{\tilde r^2+1}+2\right)}^{3/2}}
 e^{- \gamma_f^{\tilde r}\tau},\ \tau\gg 1
    \label{eq:WTDlong}
\end{equation}
for short and long times, respectively, where
\begin{equation}
\gamma_i^{\tilde r} = \frac{8 \tilde r^4+5 \tilde r^2+2}{2+2 \tilde r^2-4 \tilde r^4},
\end{equation}
and
\begin{equation}
        \gamma_f^{\tilde r} = \frac{\tilde r^2 {\left(\sqrt{\tilde r^2+1}+\tilde r \right)}}{2 {\left(\tilde r^2+\tilde r \sqrt{\tilde r^2+1} +\sqrt{\tilde r^2+1}+\tilde r +1\right)}}
\end{equation}
are the corresponding decay rates, which are indicated by dashed lines in Fig.~\ref{fig:wt}(a). The decay rate at long times is bounded as $\gamma_f^{\tilde r} \leq (\sqrt{2}-1)/2\simeq 0.21$, while the initial decay rate is only bounded from below as $1\leq \gamma_i^{\tilde r}$.

By contrast, for the coherent drive, the waiting time distribution at $\bar n=0$ reduces to the simple expression
\begin{equation}
    \begin{split}
        \mathcal{W}_\Omega (\tau) = \tilde\Omega^2 e^{-\tilde\Omega^2 \tau},
    \end{split}
\end{equation}
corresponding to a Poisson process as seen in Fig.~\ref{fig:wt}(b). At finite temperatures, the decay of the waiting time distribution at long times is governed by the rate
\begin{equation}
    \gamma_f^\Omega =  \frac{\tilde\Omega^2(\bar n+1)}{1+4\bar n (\bar n +1)}
    \label{eq:longt_decay}
\end{equation}
as illustrated with a dashed line in Fig.~\ref{fig:wt}(b).

The waiting time distribution at zero time delay is interesting since it yields the average number of photons in the cavity right after a photon emission, $\bar n(0)/\bar{n} =\langle \tau\rangle \mathcal{W}(0)$. Without the drive, we have $\bar n(0)/\bar{n} = 2$, meaning that \revision{the average number of photons in the cavity right after a photon emission is twice as large as the steady-state average}, reflecting the bunching of the photons~\cite{brange2019photon}. For the parametric drive, we find
\begin{equation}
    \label{eq:gainamplifier}
    \frac{\bar{n}_r(0)}{\bar{n}_r} =\frac{8 \bar n^2+(4\bar n (\bar n+3)+1) \tilde  r^2+2 \tilde r^4}{\left(2 \bar n+\tilde r^2\right)^2},
\end{equation}
which displays an interesting, non-monotonous dependence on the driving strength $\tilde r$ and the temperature through $\bar n$ as shown in Fig.~\ref{fig:gain}(a). For a given temperature, the largest value of this ratio is reached by the relation $\bar n=\tilde r^2/2$ and inserting this expression into Eq.~(\ref{eq:gainamplifier}), we obtain the dashed line in Fig.~\ref{fig:gain}(a). This relation implies that, at low temperatures $\bar{n} \ll 1$, the photon number within the cavity is strongly increased by an emission as $\bar{n}_r(0)/\bar{n}_r \simeq \frac{1}{16}(1/\bar{n} + 40)$. We also show the relation in Fig.~\ref{fig:gain}(b) and note that for $\tilde r=1$, the ratio is independent of $\bar n$ and obtains the value~3 even though $\bar{n}_r$ diverges.

For the coherent drive, the average number of cavity photons after an emission decreases monotonously with increasing coupling, and we find the simpler expression 
\begin{equation}
    \label{eq:gaindrive}
        \frac{\bar{n}_\Omega(0)}{\bar{n}_\Omega} =  \frac{2 \bar n^2+4 \bar n \tilde \Omega ^2+\tilde \Omega ^4}{(\bar n+\tilde \Omega ^2)^2},
\end{equation}
which takes values between one and two. Here, the value of one corresponds to the zero-temperature limit, which is governed by the coherent drive, while the value of two is reached at high temperatures, where thermal effects dominate.
At zero temperature, the detection of a photon reveals no  information about the photons in the cavity statistics, as the photon emission process is Poissonian.

\begin{figure*}
    \centering
    \includegraphics[width=\textwidth]{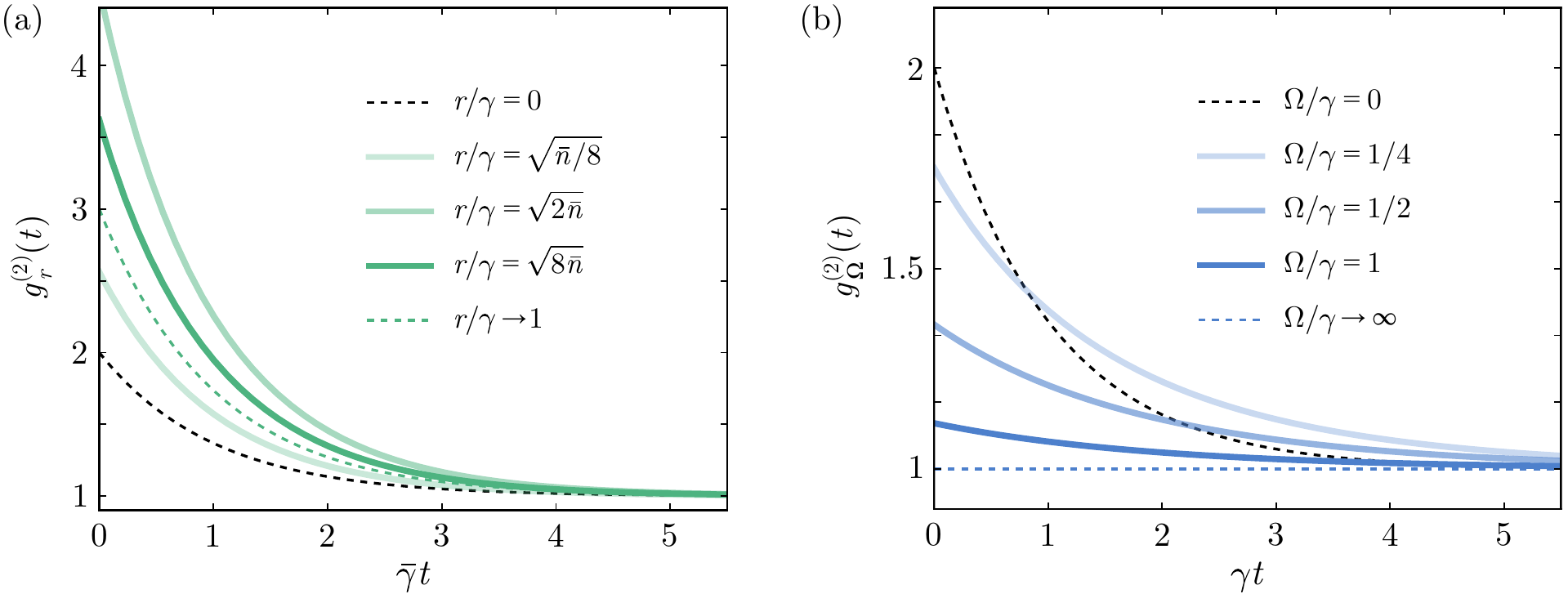}
    \caption{Second-order coherence. (a) $g^{(2)}$-functions for the parametric drive with $\bar n= 1/16$ and we have defined $\bar \gamma = \gamma(1-r/\gamma)$. (b) Similar results for the coherent drive. The black dashed lines correspond to a non-driven thermal microwave cavity.}
    \label{fig:g2}
\end{figure*}

\section{Second-order coherence}\label{ch:g2fun}
As an alternative to the waiting time distribution, we consider the second-order degree of coherence~\cite{PhysRev.130.2529,PhysRevB.85.165417}
\begin{equation}
    g^{(2)}(\tau) =\frac{\langle \hat a^\dagger(0) \hat a^\dagger(\tau) \hat a(\tau) \hat a(0) \rangle}{\langle \hat a^\dagger(\tau) \hat a(\tau) \rangle\langle \hat a^\dagger(0) \hat a(0) \rangle}.
\end{equation}
The $g^{(2)}$-function is important for determining if the emitted photons are bunched [$g^{(2)}(0)>g^{(2)}(\tau)$] or anti-bunched [$g^{(2)}(0)<g^{(2)}(\tau)$]~\cite{carmichael2009statistical}, and it is typically easier to measure compared to the waiting time distribution since it does not depend on the detector efficiency. 

The $g^{(2)}$-functions can be obtained from the generating function (see Appendix~\ref{ap:g2} for details),
and as an important check, we find the known expression~\cite{landi2023current}
\begin{equation}\label{eq:g2par}
    g^{(2)}_r(\tau) = 1+\frac{\tilde J_0^2}{\tilde J_r^2}\sum_{\nu=\pm \tilde r}
        \frac{e^{(\nu -1) | \tau|}}{2}{\left(\frac{1 + \nu/2\bar n }{1- \nu}\right)}^2 
\end{equation} 
for the parametric drive and
\begin{equation}\label{eq:g2coh}
        g^{(2)}_\Omega(\tau) = 1+ \frac{\tilde J_0^2}{\tilde J_\Omega^2}{\left(e^{-|\tau|}+ e^{-|\tau|/2}\frac{2\tilde \Omega^2}{\bar n}\right)} 
\end{equation}
for the coherently driven cavity. In Fig.~\ref{fig:g2}, we show the $g^{(2)}$-functions for different driving strengths, and we see that the photons are bunched in both cases. We note that $g^{(2)}(0)$ also yields the relative number of cavity photons right after a photon has been emitted~\cite{PhysRevA.39.1200}.

By comparing the $g^{(2)}$-function and the waiting time distribution, we can determine if the photon emissions constitute a renewal process, implying that subsequent waiting times are correlated. For a renewal process, the two are related in the Laplace domain as~\cite{PhysRevA.39.1200,Dasenbrook:2014,PhysRevB.91.195420}
\begin{equation}
\label{eq:renewal}
    g^{(2)}(s) = \langle \tau\rangle \frac{\mathcal{W}(s)}{1-\mathcal{W}(s)},
\end{equation}
where $g^{(2)}(s)$ and $\mathcal{W}(s)$ are the Laplace transformed distributions, which can be found in Appendix~\ref{ap:g2}. By checking this relation, we find that it generally does not hold for either of the drives. Only for the coherent  drive, the emission events become a renewal process at zero temperature, $\bar n=0$, where the stationary state is a coherent state, and the photon emission process is Poissonian. 

\begin{figure*}
    \centering
    \includegraphics[width= \textwidth]{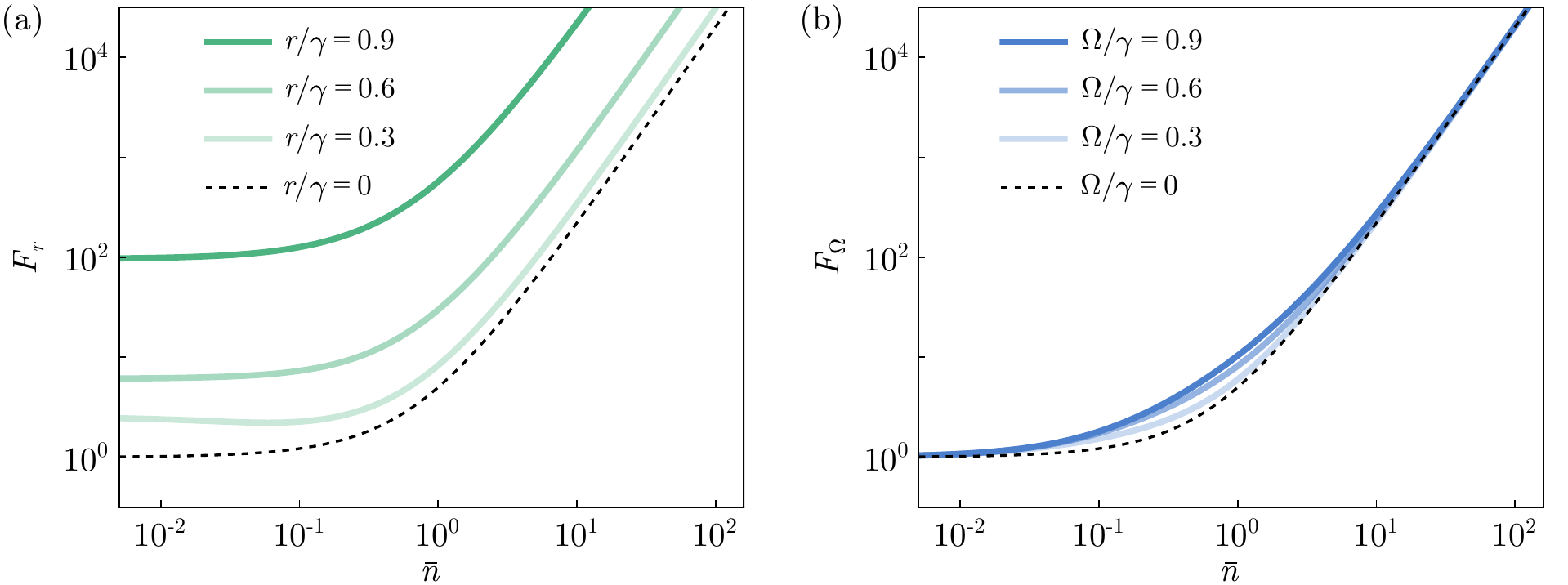}
    \caption{Fano factor of the photon current. (a) Fano factor for the parametric drive as a function of the temperature, given by $\bar n$, and with different couplings. (b) Similar results for the coherent drive. The dashed lines correspond to a non-driven cavity.}
    \label{fig:fano}
\end{figure*}

\section{Long-time limit}\label{ch:longtime}
In addition to the time-resolved quantities that we have considered so far, it is also interesting to investigate the photon emission statistics collected over a long time-duration. To this end, we consider the long-time limit of the scaled factorial cumulant generating function, 
\begin{equation}
\mathcal{F}(\zeta) = \lim_{\tau\rightarrow \infty} \ln[\mathcal M (\zeta,\tau)]/\tau,
\end{equation}
which for each of the two drives reads
\begin{equation}
        \mathcal{F}_r(\zeta) =[2-(\xi_{\tilde r}+\xi_{-\tilde r})]/4
\end{equation}
and
\begin{equation}
        \mathcal{F}_\Omega(\zeta) =\mathcal{F}_0(\zeta)+\frac{\zeta \tilde \Omega ^2 (\bar n+1) }{ 1-\zeta [4  \bar n (\bar n+1)]},
\end{equation}
respectively, where $\mathcal{F}_0 (\zeta) = \mathcal{F}_r (\zeta)\vert_{ r=0}$ is the scaled  generating function of a thermal cavity, and $\xi_\nu$ is given by Eq.~\eqref{eq:xi}. The scaled factorial cumulants are then obtained as $\tilde\kappa_m = \partial_\zeta^m\mathcal{F}(\zeta)\vert_{\zeta = 0}$, and we find
\begin{equation}\label{eq:cumulants}
    \tilde \kappa_m^r = \frac{  (\bar n+1)^m (2 m)!}{8{\left(m-\frac{1}{2}\right)}  m!}\sum_{\nu=\pm \tilde r} \frac{(\bar n+\nu /2)^m}{(1-\nu )^{2m-1}}
\end{equation}
and
\begin{equation}\
\tilde\kappa_m^\Omega =\kappa_m^0+ \tilde \Omega^2 m! (4\bar n)^{m-1}(\bar n+1)^m
\end{equation}
for each of the two drives with $\tilde\kappa_m^0 = \tilde\kappa_m^{r=0}$. Interestingly, all the factorial cumulants are positive, which appears to be a typical feature of noninteracting bosons~\cite{Beenakker:1998,Beenakker:2001,brange2019photon}. By contrast, the factorial cumulants of non-interacting electrons alternate in sign with the order~\cite{Kambly2013}.

The first factorial cumulant is the average current, which is shown in Fig.~\ref{fig:fig1}(e)-(f) for the two drives. In Fig.~\ref{fig:fano}, we plot the ratio of the current fluctuations over the average current, which is given by the Fano factor
\begin{equation}
F=1+\kappa_2/\kappa_1,
\end{equation}
and we see clear differences between the two drives. The parametric drive produces a higher level of noise at all temperatures compared to the coherent drive, which is dominated by thermal fluctuations, in particular at high temperatures. At low temperatures, the photon emission statistics from the coherently-driven cavity becomes Poissonian, and the Fano factor equals one for all values of the coupling $ \Omega$. Only in the region around $\bar n\simeq 1$, the Fano factor clearly depends on the coupling.

\section{Large-deviation statistics}\label{ch:largedev}
Finally, we consider the large-deviation statistics of the photon current. To this end, we formally write the probability distribution for the photon emission statistics by inverting the generating function as
 \begin{equation}
  P(n,t)=\frac{1}{2\pi}\int_{-\pi}^\pi d\chi G(e^{i\chi},t)e^{-in\chi}.
 \end{equation}
At long times, we may express the distribution as
\begin{equation}
P(J=n/t,t)=\frac{1}{2\pi}\int_{-\pi}^\pi d\chi e^{[\mathcal{F}(\chi)-iJ\chi] t},
\end{equation}
where $J=n/t$ is the photon emission current, and $\mathcal{F}(\chi)$ is the scaled factorial cumulant generating function with the substitution, $\zeta \rightarrow e^{i\chi}-1$. In this form, the integral is amenable to a saddle-point approximation, which allows us to express the large-deviation statistics as 
 \begin{equation}
     \frac{\ln[P(J,t)]}{t} \simeq  \mathcal{F}(\chi_0)-iJ\chi_0,
 \end{equation}
where the imaginary saddle-point, $\chi_0$, solves the saddle-point equation, $\mathcal{F}'(\chi_0)=iJ$.

\begin{figure*}
    \centering
    \includegraphics[width=\textwidth]{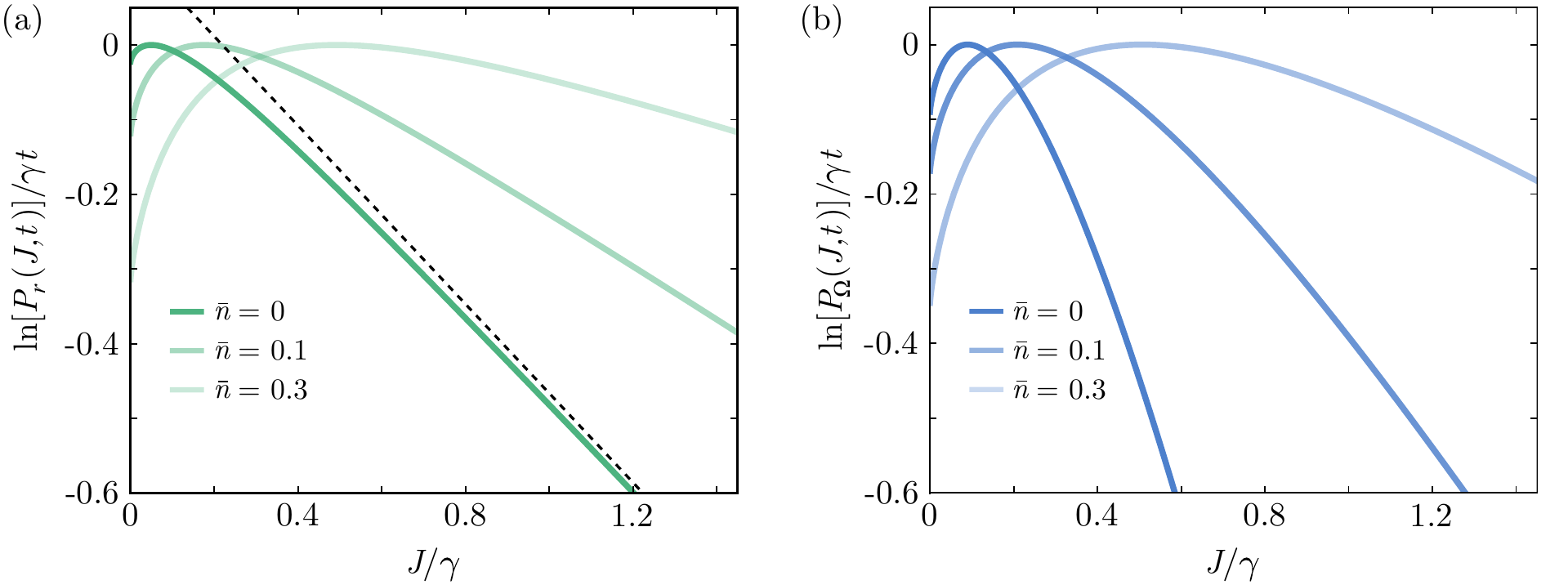}
    \caption{Large-deviation statistics of the photon emission current. (a) Statistics of the photon current for the parametric drive with the coupling $r = 0.3\gamma$ and different temperatures. The dashed line is the analytic expression in Eq.~(\ref{eq:ldfapp}). (b)  Statistics of the photon current for the coherent drive with the coupling $\Omega = 0.3\gamma$ and different temperatures.}
    \label{fig:largedev}
\end{figure*}

In Fig.~\ref{fig:largedev}, we show the large-deviation statistics for the two drives at different temperatures. At large temperatures, the distributions are similar as they are both dominated by thermal effects. On the other hand, clear differences become visible as the temperature is lowered. In particular, at zero temperature, we can find simple expressions for the large-deviation statistics.
For the parametric drive with $\bar n =0$ , we find for large currents
\begin{equation}
    \frac{\ln[P_r(J,t)]}{ \gamma t} \simeq \frac{1}{2}-\frac{1}{4} \sqrt{2\tilde r^2+2}-\frac{\tilde J}{2} \ln \left[\frac{{\left(\tilde r^2+1\right)}^2}{4 \tilde r^2}\right],
    \label{eq:ldfapp}
\end{equation}
which predicts a linear dependence for $J\gg \gamma$ as shown with a dashed line in Fig.~\ref{fig:largedev}(a). \revision{In this context, we note that the statistics of radiation emitted at a Josephson parametric resonance was~explored~in~Ref.~\cite{Padurariu:2012}.}
For the coherent drive at zero temperature, we find
 \begin{equation}
     \frac{\ln[P_\Omega(J,t)]}{\gamma t } = \tilde J-\tilde \Omega ^2-\tilde J \ln {\left(\frac{\tilde J}{\tilde \Omega ^2}\right)},
 \end{equation}
 which is the statistics of a Poisson process.

\section{Conclusions}
\label{sec:conc}
We have investigated the statistics of photons that are emitted from a driven microcavity and explored the differences between a parametric drive and a coherent drive.
To this end, we have found the generating function for the photon emission statistics using a technique based on Gaussian states and the solution of a Riccati equation. For both drives, the photon emission statistics can be understood in terms of two independent processes. We have calculated the distribution of waiting times between subsequent photon emissions and identified important differences between the two drives. Moreover, from the $g^{(2)}$-function of the outgoing photons, we have shown that the photon emissions do not constitute a renewal process since subsequent waiting time are correlated. In the long-time limit, we have obtained simple expressions for the Fano factor and the factorial cumulants, which are all positive. We have also found marked differences between the two drives in the large-deviation statistics of the photon current. Throughout the paper, we have focused on finite-temperature effects, which are of particular importance for microwave photons, both for calorimetric detection schemes as well as in the broader framework of quantum thermodynamics. 

Our work can be extended in many directions. For example, it is possible to include several thermal reservoirs and investigate the statistics of the heat that will flow through the cavity due to a temperature difference between them. Such investigations may be of relevance in the context of thermodynamic uncertainty relations for quantum systems. Within our formalism, it is also possible to describe several coupled microwave cavities with the potential of generating non-classical correlations and entanglement between the outgoing photons. As single-photon detectors in the gigahertz regime are currently being developed, our predictions for the photon emissions statistics may be observable in future experiments.

\begin{acknowledgments}
We thank D.~Larsson Persson for fruitful discussions and acknowledge the support from the Research Council of Finland through the Finnish Centre of Excellence in Quantum Technology (Grant No.~352925) and Grant No.~331737. This work was partially supported by the Wallenberg Centre for Quantum Technology (WACQT) funded by Knut and Alice Wallenberg Foundation.
\end{acknowledgments}

\appendix
\section{Gaussian states}\label{ap:gauss}
The driven cavity in the main text is naturally described by Gaussian states. Here, we establish their definition and the relevant notation, which are then used in the following Appendix to develop an approach to the full counting statistics of driven cavities. 

The characteristic function of a state $\hat\rho$ is given by
\begin{equation}
	\chi_{\hat\rho}(\boldsymbol \alpha)=\mathrm{tr}\{\hat\rho 
    \exp{\left( \hat{a}^\dagger\alpha - \hat a \alpha^*\right)}\} 
\end{equation}
where $\alpha$ is a complex variable and $\boldsymbol \alpha = (\alpha,\ \alpha^*)^T$. \revision{The Wigner function of the state is then given by the Fourier transform of the characteristic function as
\begin{equation}
W(x,p) = \frac{1}{2\pi^2}\int d^2 \alpha\,\chi_{\hat \rho}(\boldsymbol\alpha) e^{ i \frac{\alpha + \alpha^*}{\sqrt{2}} \tilde p + \frac{\alpha - \alpha^*}{\sqrt{2}} \tilde x}, 
\end{equation}
where $\tilde x=x/x_0$ and $\tilde p= x_0 p/\hbar$ are given in terms of the usual oscillator length~$x_0$.} Creation and annihilation operators acting on a density matrix produce the following characteristic functions~\cite{carmichael2009statistical},
\begin{equation}\label{eq:rules}
\begin{split}
\chi_{\hat a\hat\rho}(\boldsymbol\alpha)=(-\partial_\alpha^*-\alpha/2) \chi_{\hat\rho}(\boldsymbol\alpha),\\
    \chi_{\hat\rho \hat a}(\boldsymbol\alpha)=(-\partial_\alpha^*+\alpha/2)\chi_{\hat\rho}(\boldsymbol\alpha),\\
    \chi_{\hat{a}^\dagger\hat\rho}(\boldsymbol\alpha)=(\partial_\alpha-\alpha^*/2) \chi_{\hat\rho}(\boldsymbol\alpha), \\
    \chi_{\hat\rho \hat{a}^\dagger}(\boldsymbol\alpha)=(\partial_\alpha+\alpha^*/2) \chi_{\hat\rho}(\boldsymbol\alpha).
\end{split}
\end{equation}
A Gaussian state has the characteristic function~\cite{Adesso}
\begin{equation}
\chi_{\hat\rho}(\boldsymbol\alpha ) = e^{- \boldsymbol\alpha ^\dagger \sigma_z \Theta \sigma_z \boldsymbol\alpha/2+ \boldsymbol\alpha^\dagger \sigma_z \boldsymbol d},
\end{equation}
where $\sigma_i$ are the Pauli matrices, and we refer to $\boldsymbol d = (d,\ d^*)^T$ and $\Theta$ as the displacement vector and the covariance matrix, respectively, which are related to the first and second moments of~$\hat a$.
Specifically, they read
\begin{equation}
        \boldsymbol d = (\langle \hat a\rangle,\ \langle \hat{a}^\dagger\rangle)^T
\end{equation}
and
\begin{equation}
\Theta =\begin{pmatrix}
        \langle \delta \hat{a}^\dagger \delta \hat a \rangle & \langle \delta \hat{a}^\dagger \delta \hat{a}^\dagger \rangle\\
        \langle \delta \hat a \delta \hat a \rangle & \langle \delta \hat{a}^\dagger \delta \hat a \rangle
        \end{pmatrix}+\mathds 1/2
\end{equation}
where $\delta \hat a = \hat a - \langle \hat a \rangle$ is the deviation from the average value, and $\mathds{1}$ is the identity matrix. The two stationary states in the main text, which are squeezed or displaced thermal states, are both examples of Gaussian states.

\section{Generating function}\label{ap:MGF}
To find the generating function of the photon emission statistics, we assume that the density matrix including the counting field can be described as a Gaussian state~\cite{kerremans2022probabilistically}. As we will see, this ansatz allows us to find the unique solution of the Lindblad equation including the counting field, demonstrating that the ansatz indeed is correct. Thus, we consider the characteristic function
\begin{equation}
    \chi_{\hat\rho(\zeta,t)} (\boldsymbol\alpha ) = \mathcal{M}(\zeta,t) e^{- \boldsymbol\alpha ^\dagger \sigma_z \Theta(t) \sigma_z \boldsymbol\alpha/2+ \boldsymbol\alpha^\dagger \sigma_z \boldsymbol d(t)},
\end{equation}
where we have used that it yields the trace of the density matrix at $\boldsymbol\alpha=0$, such that $\chi_{\hat\rho(\zeta,t)} (0 )=\mathrm{tr}\{\hat\rho(\zeta,t)\}=\mathcal{M}(\zeta,t)=\exp[\mathcal C (\zeta,t)] $. Taking the derivative of this characteristic function with respect to time, we get 
\begin{equation}
\begin{split}\label{eq:charderivation}
{\left(\frac{d}{dt} \mathcal{C}- \boldsymbol\alpha ^\dagger \sigma_z \frac{d\Theta}{dt} \sigma_z \boldsymbol\alpha/2 + \boldsymbol\alpha^\dagger \sigma_z \frac{d }{dt}\boldsymbol d\right)}  \chi_{\hat\rho} =&\chi_{\mathcal L_\zeta \hat\rho},
\end{split}
\end{equation}
where the time arguments of $\mathcal C$, $\boldsymbol d$, and $\Theta$  have been omitted to simplify the notation. Using  the Liouvillean including the counting field in Eq.~(\ref{eq:lindbladcf}) in combination with the expressions in Eq.~\eqref{eq:rules} on the right hand side of Eq.~\eqref{eq:charderivation}, we then obtain equations of motion for the displacement vector and the covariance matrix.

For the parametric drive, we now find
\begin{equation}
\boldsymbol d (t) = \boldsymbol 0,
\end{equation}
having used the initial condition $\boldsymbol d (0) = \boldsymbol 0$ in the stationary state, together with the equations of motion
 \begin{equation}
 \frac{d}{dt}\Theta(t)= \Theta(t) X\Theta(t)+W \Theta(t)+\Theta(t) W^{\dagger } +F,
 \end{equation}
 and
\begin{equation}
\label{eq:CGF}
 \frac{d}{dt}\mathcal C(\zeta,t)  = \mathrm{tr}\{ (\Theta(t)-\mathds 1/2) X\} /2 ,
\end{equation}
where we have defined the matrices
\begin{equation}
\begin{split}
        X & = \gamma \zeta (\bar n+1)  \mathds 1,\\
        F & = \gamma (\bar n+1/2) \mathds 1 + X/4,\\
        W & = -\gamma(\tilde r e^{-i 2 \phi \sigma_z}i \sigma_z\sigma_x+ \mathds 1)/2- X/2.
\end{split}
\end{equation} 
The initial condition for the covariance matrix, $\Theta(t=0)\equiv\Theta_0$, is found from the stationary state of the Lindblad equation without the counting field as
\begin{equation}
\begin{split}
    (W \Theta_0+\Theta_0 & W^\dagger + F)\vert_{\zeta =0} = 0,
    \end{split}
\end{equation}
since the stationary state has been reached as we start counting photons, which also implies that $C(\zeta,t=0)=0$. The equation of motion for the covariance matrix~$\Theta$ with the counting field~$\zeta$ is known in control theory as a Riccati equation, and it can be solved analytically~\cite{abou2012matrix}.

First, we find the covariance matrix~$\Theta_\infty$ at long times. To this end, we introduce the symplectic matrix
\begin{equation}
    \mathcal H = \begin{pmatrix}
    W^\dagger & X\\
    -F & -W
    \end{pmatrix},
\end{equation}
which has four eigenvectors. We choose two of them, $v_1$ and $v_2$, so that the  covariance matrix~$\Theta_\infty$ yields the stationary solution~$\Theta_0$ at $\zeta=0$. Specifically, by constructing the $2\times 4$ matrix  $C = (v_1, v_2)$ and defining two $2\times 2$ matrices so that $C = (C_1,C_2)^T$, we eventually find
$\Theta_\infty = C_2C_1^{-1}$.  

Once we have the covariance matrix at long times, we obtain the full time-dependent solution from the
matrix 
\begin{equation}
L(t) = [\Theta(t)-\Theta_\infty]^{-1},
\end{equation}
which evolves according to the Lyapunov equation
\begin{equation}
    \frac{d }{dt}L = -(W+ \Theta_\infty X) L - L (W+ \Theta_\infty X)^\dagger -X.
\end{equation}
Since this a linear equation for $L$, it can be integrated  and solved by first vectorizing $L$. We then find $\Theta(t)$  by inverting~$L(t)$. As the last step, we obtain the generating function by inserting the covariance matrix into Eq.~(\ref{eq:CGF}), and after some algebra we arrive at $\mathcal M_r(\zeta,t)$ in Eq.~\eqref{eq:MGFr}.

The coherently driven cavity is characterized by the occupation $\bar n_\zeta(t)$ and the displacement $d (t) = \langle \hat a \rangle$ with the initial conditions $\bar n_\zeta(0) =  \bar n$ and $d(0) = i \Omega e^{-i\phi_\Omega}$. The equations of motions for each of these quantities can be obtained just as above, and they read
\begin{equation}
    \frac{d }{dt}\bar n_\zeta (t) = \gamma \big(\bar n  -  \bar n_\zeta(t) + \zeta (1 +\bar n)  \bar n^2_\zeta(t)\big),
\end{equation} 
together with
\begin{equation}
    \frac{d}{dt}d(t)=\gamma \Big(i e^{-i \phi_\Omega} \tilde \Omega - d(t)\big(1  +(\xi_0 ^2-1) \bar n_\zeta(t)/2 \bar n\big)\Big)/2,
    \end{equation} 
and
\begin{equation}
\frac{d}{dt} \mathcal C (\zeta,t) =\gamma  (\xi_0^2-1) (\bar n_\zeta(t)+|d(t)|^2)/4 \bar n  
\end{equation}
Solving these equations, we then arrive at Eq.~\eqref{eq:MGFdrive}.

\section{Second-order coherence
}\label{ap:g2}
The $g^{(2)}$-function can be related to the noise as
\begin{equation}
    g^{(2)}(\tau) = 1+ \int_{-\infty}^\infty d \omega e^{-i \omega \tau} (F(\omega) -1)/2\pi J,
\end{equation}
where $F(\omega)=S(\omega)/J$ is the Fano factor, and we can find the noise using MacDonald's formula~\cite{0034-4885-12-1-304,FLINDT2005411,PhysRevB.75.045340}, \begin{equation}
    S(\omega) = \omega \int_{0}^\infty d t \sin(\omega t) \frac{d}{d t} (\kappa_2 + \kappa_1),
\end{equation}
expressed in terms of the first and second factorial cumulants. We then find the $g^{(2)}$-functions in Eqs.~(\ref{eq:g2par})~and~(\ref{eq:g2coh}). Moreover, in the Laplace space, 
\begin{equation}
    g^{(2)}(s) = \int_0^\infty d \tau e^{-s \tau} g^{(2)} (\tau),
\end{equation}
we have
\begin{equation}\label{eq:g2s}
g^{(2)}_r(s) = 1+\frac{\tilde J_0^2}{\tilde J_r^2}\sum_{\nu=\pm \tilde r}
        \frac{(1 -\nu/2\bar n)^2}{2(1+\nu)^2(s+1+\nu )}
        \end{equation}
for the parametrically driven cavity and
\begin{equation}
        g^{(2)}_\Omega(s) = 1+ \frac{\tilde J_0^2}{\tilde 
 J_\Omega^2}{\left(\frac{1}{s+1}+ \frac{2}{2 s+1}\frac{2 \tilde \Omega^2}{\bar n}\right)}
\end{equation}
for the coherently driven cavity. In combination with the waiting time distributions, we use these expressions to check the renewal assumption in Eq.~(\ref{eq:renewal}).


%

\end{document}